\documentclass[prl,twocolumn,final,superscriptaddress,showpacs]{revtex4}

\usepackage{graphicx}
\usepackage{dcolumn}
\usepackage{bm}
\usepackage[utf8x]{inputenc}
\usepackage{srctex}
\usepackage{amsmath}
\usepackage{amsfonts}
\usepackage{amssymb}
\usepackage{graphicx}
\usepackage{xcolor}

\begin{document}

\title{Emergence of collective motion in a model of interacting Brownian particles}

\author{Victor Dossetti}
\email{dossetti@ifuap.buap.mx}
\thanks{author to whom correspondence should be addressed.}
\affiliation{Instituto de F\'isica, Benem\'erita Universidad Aut\'onoma de Puebla, Apdo.\ Postal J-48, Puebla 72570, Mexico}
\affiliation{Consortium of the Americas for Interdisciplinary Science, University of New Mexico, Albuquerque, NM 87131, USA}

\author{Francisco J.\ Sevilla}
\email{fjsevilla@fisica.unam.mx}
\affiliation{Instituto de F\'isica, Universidad Nacional Aut\'onoma de M\'exico, Apdo.\ Postal 20-364, 01000, M\'exico D.F., Mexico}

\date{\today}

\begin{abstract}

By studying a system of Brownian particles, interacting only through a local social-like force (velocity alignment), we show that self-propulsion is not a necessary feature for the flocking transition to take place as long as underdamped particle dynamics can be guaranteed. Moreover, the system transits from stationary phases close to thermal equilibrium, with no net flux of particles, to far-from-equilibrium ones exhibiting collective motion, long-range order and giant number fluctuations, features typically associated to ordered phases of models where self-propulsion is considered.

\end{abstract}


\maketitle

Collective motion is an ubiquitous phenomenon in biological groups such as flocks of birds, schools of fishes, swarms of insects, etc. The study of these far-from-equilibrium systems has attracted great interest over the last few decades, as the spontaneous emergence of such ordered phases and coordinated behavior, arising from local interactions, cannot be accounted for by the standard theorems of statistical mechanics \cite{vic12}.

Introduced almost twenty years ago, the seminal model by Vicsek \emph{et al.}\ \cite{vic95} provided a simple tool to study the transition to collective motion, in a non-equilibrium si\-tu\-ation, taking into account two basic ingredients: the \emph{self-propulsive} character of the particles and a local velocity-\emph{alignment} interaction among them. Due to the discrete nature of the model, a given particle instantaneously orients its direction of motion along the ave\-rage direction of motion of its neighbors within a radius $R$, while stochastic perturbations are considered by adding a random angle (\emph{\textquotedblleft noise\textquotedblright}) to this direction. In two dimensions, the system displays an ordered phase characterized by collective motion with \emph{long-range order} and giant number fluctuations \cite{ton05}, just below a critical value of the noise intensity that depends on the particle density. It is in this respect that this model can be considered as a paradigm of non-equilibrium phase transitions, since such ordered phases are forbidden in equilibrium (for Heisenberg-like models) by the Mermin-Wagner-Hohenberg theorem \cite{mer66}. Above this value, the ordered phase breaks down into a stationary, disordered and out-of-equilibrium one. 

Over the years, many generalizations of the model have been developed, keeping self-propulsion as an essential ingredient for the phase transition to take place, in combination with interactions of the \hbox{\textquotedblleft social\textquotedblright} type among the particles such as velocity-alignment \cite{gre04, gin10}. This has led to the development of sophisticated nonlinear friction terms \cite{erd02, dos09}, a bias that may be justified arguing, on the one hand, that the concept of self-propelled (or \emph{active} \hbox{Brownian}) particles captures the natural ability (seen in many biological systems) for the agents to develop motion by themselves \cite{sch07, gro12} and, on the other, that it is an important ingredient for pattern formation in  models of collective motion \cite{shi96}. 

In this Letter, we study the emergence of collective motion in a two-dimensional system of $N$ \hbox{\emph{passive}} \hbox{Brownian} (not-self-propelled) particles that interact among themselves only through local forces of the social type (velocity-alignment in this case). In such a way, we show that the symmetry breaking of the disordered phase corres\-ponds to the breakdown of a close-to-equilibrium state for a particular value of the ratio between two characteristic time-scales in the system: one related to the mean collision time of Brownian particles immersed in a thermal bath, and the other related to the rate of alignment among particles. Our results show that local alignment interactions suffice for the development of collective motion, with true long-range order, whenever underdamped particles (for which inertial motion cannot be neglected) are considered. Through standard numerical measurements, the far-from-equilibrium nature of these states is stablished, as they exhibit typical features of this kind of phases such as giant number fluctuations among others.

Our model is described in terms of generic stochastic differential equations for Brownian particles restricted to move within a box of linear size $\mathcal{L}$ with periodic boundary conditions, \emph{i.e.},
\begin{equation}
m\frac{d\boldsymbol{v}_{i}}{dt} = \boldsymbol{\mathcal{F}}_{i}-\gamma \boldsymbol{v}_{i}+\boldsymbol{\xi}_{i},
\label{model_eqs}
\end{equation}
where $\boldsymbol{v}_{i}=d\boldsymbol{x}_{i}/dt$ and $m$ is the mass of the particles. The last two terms on the right hand side correspond to the linear-dissipative and fluctuating forces that appear in the Langevin description of Brownian motion, respectively. The components of the vector $\boldsymbol{\xi}_{i}$ are uncorrelated Gaussian white noises with zero mean and, because of the fluctuation-dissipation relation (FDR), with auto\-corre\-lation function $\langle\xi_{i,\mu}(t)\xi_{j,\nu}(t')\rangle = 2\gamma k_{B}T \delta_{i,j} \delta_{\mu,\nu} \delta(t-t')$, where $\xi_{i,\eta}$ is the $\eta$-th Cartesian component of $\boldsymbol{\xi}_{i}$, $k_{B}$ is the Boltzmann constant, $T$ the temperature of the bath, and $\delta_{u,w}$ and $\delta(\tau)$ are the Kronecker delta and the Dirac delta function, respectively.

The alignment behavior is implemented through the force $\boldsymbol{\mathcal{F}}_{i} = \Gamma \left[\boldsymbol{f}_{i} - \hat{\boldsymbol{v}}_{i}\left(\boldsymbol{f}_{i}\cdot\hat{\boldsymbol{v}}_{i}\right)\right]$, that corresponds to the two-dimensional form of $\Gamma \left[\hat{\boldsymbol{v}}_{i} \times (\boldsymbol{f}_{i} \times \hat{\boldsymbol{v}}_{i})\right]$, with $\hat{\boldsymbol{v}}_{i}$ being the unitary vector in the direction of $\boldsymbol{v}_{i}$ and $v_{i}=\vert\boldsymbol{v}_{i}\vert$, while $\boldsymbol{f}_{i}=\left[N_{R}(i)\right]^{-1}\sum_{j\in\Omega_{R}(\boldsymbol{x}_{i})}\hat{\boldsymbol{v}}_{j}$ corresponds to the local ins\-tan\-ta\-neous direction of motion, \emph{i.e.}, the arithmetic average of the direction of motion of the $N_{R}(i)$ particles that surround the $i$-th particle within a neighborhood $\Omega_{R}(\boldsymbol{x}_{i})$ of radius $R$. The coupling factor $\Gamma$ is a measure of how fast the velocity vector of a single particle aligns along the direction of $\boldsymbol{f}_{i}$. It is easy to check its non-propelling character since \hbox{$\boldsymbol{\mathcal{F}}_{i}\cdot\hat{\boldsymbol{v}}_{i}=0$}. The alignment interaction considered here corresponds to a simple generalization, with a finite aligning rate, of the one introduced in \cite{vic95}. We must mention that alignment interactions with finite aligning rates have been considered before in some other forms \cite{dos09, gro12, mog96}.

We choose as time, speed and length scales the quantities: $\tau_{0} = m/\gamma$, $v_{0}=\sqrt{2 k_{B}T /m}$ and $r_{0} = v_{0} \tau_{0}$, respectively. In this way, the number of independent parameters in our model is reduced to three: the dimensionless alignment-coupling constant $\tilde{\Gamma} = v_0^{-1} \gamma^{-1} \Gamma$, the dimensionless interaction range $R/r_{0}$, and the dimensionless particle density $\rho = N/L^{2}$ with $L = \mathcal{L}/r_{0}$. Without loss of generality, we fix $R/r_{0} = 1$ further on. All numerical results presented here, were obtained by integrating the set of Eqs.\ (\ref{model_eqs}) with a modified version of the velocity-Verlet algorithm \cite{gro97} with an integration time-step $\Delta t = 0.01$. Random initial conditions where taken for the position of the particles while their initial velocities where drawn from the equilibrium Maxwell-Boltzmann velocity distribution.

\begin{figure}[t]
\includegraphics[width=0.47\textwidth]{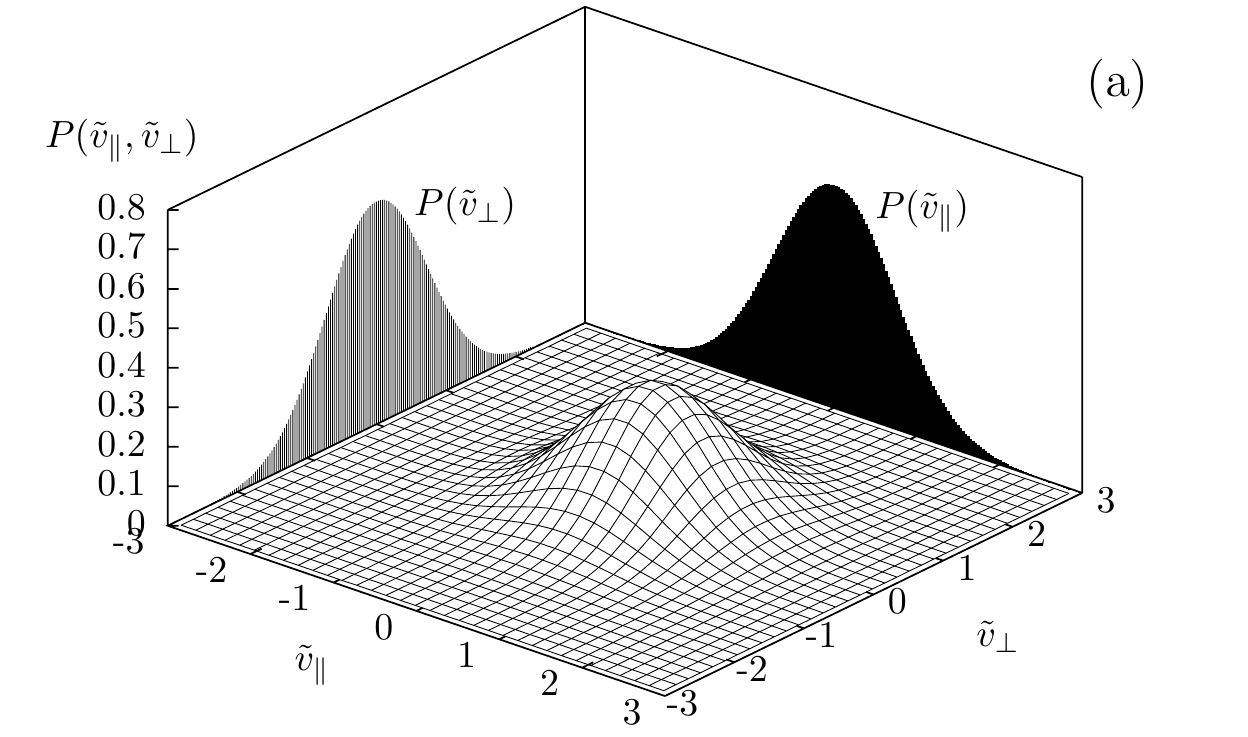}
\includegraphics[width=0.47\textwidth]{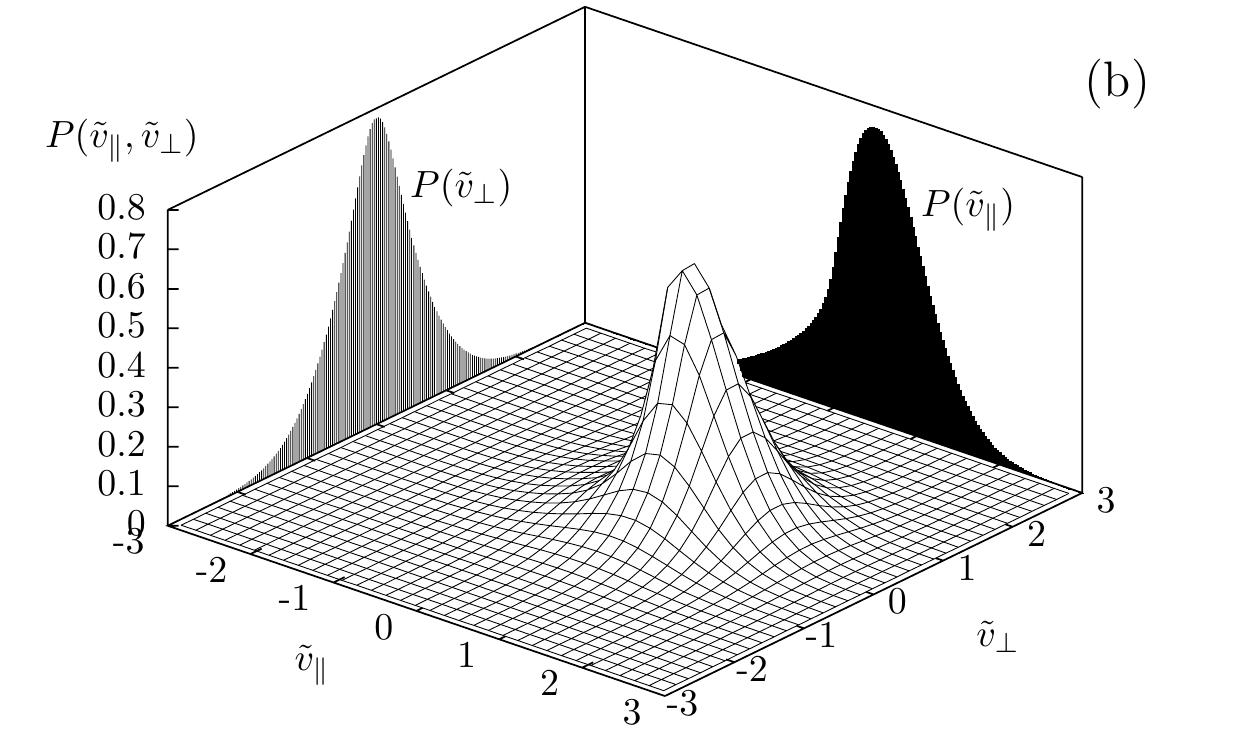}
\caption{Stationary probability distribution $P(\tilde{v}_{\|},\tilde{v}_{\bot})$ of the individual velocity of the particles, $\tilde{\boldsymbol{v}}_i$, projected along the direction of the mean velocity of the group, $\tilde{v}_{\|}$, and in the transversal direction, $\tilde{v}_{\bot}$, for a subcritical $\tilde{\Gamma}=1$ in (a) and a supercritical $\tilde{\Gamma}=8$ in (b) for systems with $\rho = 1$ and $L = 96$. The distributions shown in the vertical planes correspond to the integral of $P(\tilde{v}_{\|},\tilde{v}_{\bot})$  over $\tilde{v}_{\|}$ and $\tilde{v}_{\bot}$, yielding $P(\tilde{v}_{\bot})$ and $P(\tilde{v}_{\|})$, respectively. Notice the asymmetry of $P(\tilde{v}_{\|})$ in (b) for the ordered state with $\tilde{\Gamma}=8$, which is the only distribution not centered around zero.}
\label{fig1}
\end{figure}

In the absence of interactions ($\tilde{\Gamma} = 0$), the particle dyna\-mics is constrained by the FDR as in the standard description of Brownian motion, therefore, the stationary-state distribution of the single particle velo\-ci\-ties corresponds to that of equilibrium with the highest rotational symmetry. Interestingly, for small values of $\tilde{\Gamma}$, where the alignment time-scale is smaller than that related to the FDR, the system remains in a disordered phase still characterized by properties of \emph{thermal equilibrium} with no net transport, normal diffusion of individual particles and Maxwellian probability densities of single particle velocities as shown in Fig.\ \ref{fig1}(a).

\begin{figure}
\includegraphics[width=0.49\textwidth]{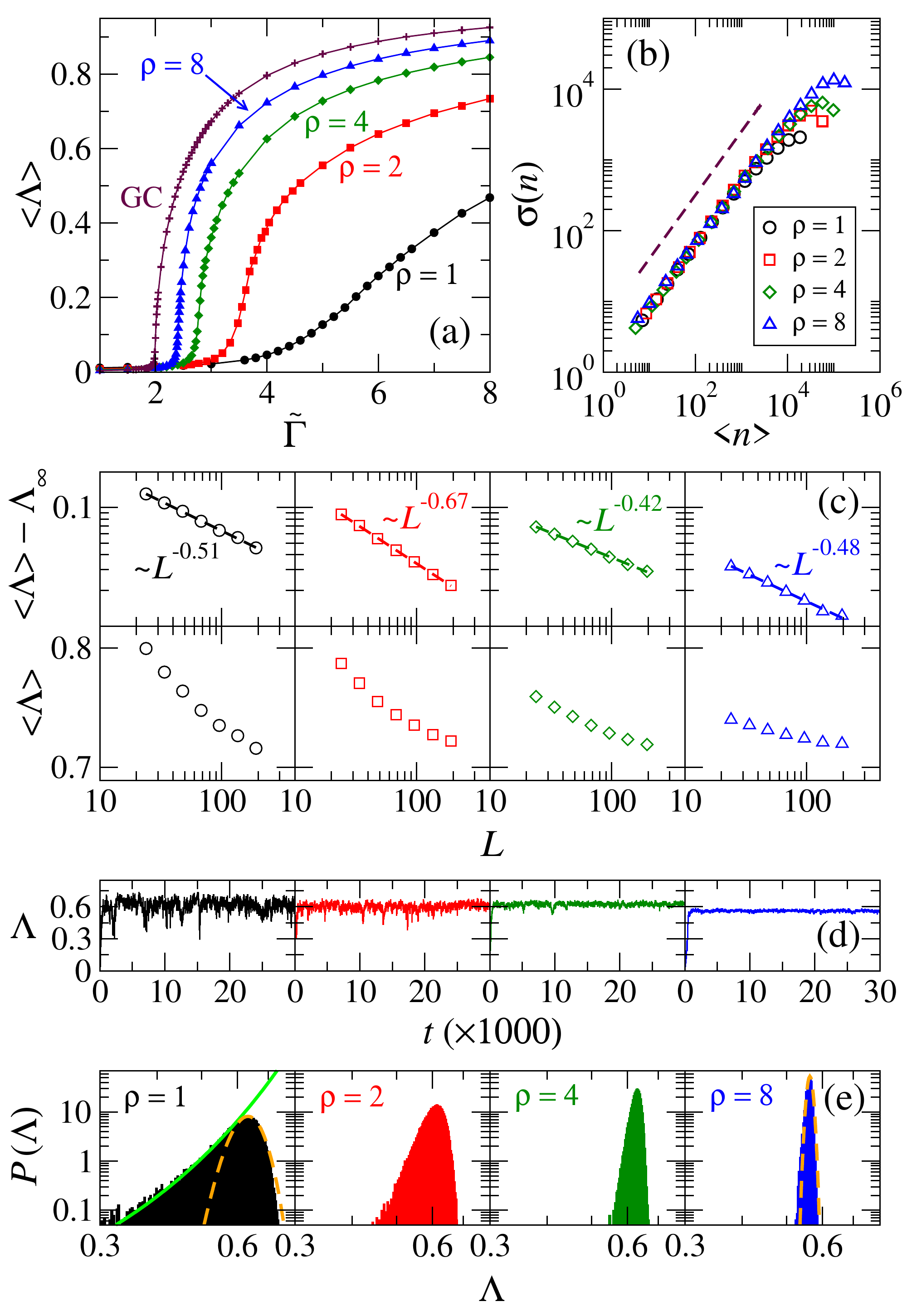}
\caption{(Color online) (a) Stationary order parameter $\langle \Lambda \rangle$ vs $\tilde{\Gamma}$ for systems with $L=96$ and diffe\-rent density values. The curve with plus symbols corresponds to the globally-coupled (GC) case for a system with $\rho=1$ and $L=320$. \hbox{(b) Giant} density fluctuations, $\sigma(n)$ vs $\langle n \rangle$, for ordered phases with \hbox{$\langle \Lambda \rangle \approx 0.7$} of systems with $L=196$. The dashed line has a slope of 0.9. (c) In the bottom row, log-log plots of $\langle\Lambda\rangle$ vs $L$ for systems with $\rho = 1, 2, 4, 8$ and $\tilde{\Gamma} = 18, 8, 5, 4$ from left to right. The log-log plots of the top row show the same data from which $\Lambda_{\infty} = 0.669538, 0.698997, 0.689226, 0.706573$ has been respectively subtracted. The dashed lines corres\-pond to power-law-decay fits, with fitting error smaller than 2\%. (d) $\Lambda$ vs $t$ for ordered phases with $\langle \Lambda \rangle \approx 0.6$ of systems with $L=96$: from left to right $\rho=1,2,4,8$. (e) Log-log plot of the stationary $P(\Lambda)$ vs $\Lambda$ for the systems in (d). The solid green curve corresponds to an exponential fit while the orange dashed curves to Gaussian fits.}
\label{fig2}
\end{figure}

As $\tilde{\Gamma}$ increases above a critical value, the system develops a self-induced far-from-equilibrium phase that exhi\-bits a net flux of particles and non-Maxwellian probability densities of single particle velocities as shown in Fig.\ \ref{fig1}(b). We characterize the ordered collective motion in this phase through the accumulated order para\-me\-ter $\langle\Lambda\rangle=\lim_{T\rightarrow\infty}T^{-1}\int_{0}^{T}\Lambda(t)\, dt$, that is calculated from the instantaneous  $\Lambda(t) = |N^{-1} \sum_{i=1}^{N} \, \exp[i\theta_{i}(t)]|$ with $\theta_{i}(t)$ defined from $\boldsymbol{v}_{i} = v_i \exp[i\theta_i(t)]$. The critical point $\tilde{\Gamma}_c$ that separates the disordered and ordered phases decreases with $\rho$, reaching a limit value when $\rho \rightarrow \infty$, that corresponds to that of the globally-coupled (GC) regime ($R/\mathcal{L} = 1$) [Fig.\ \ref{fig2}(a)]. This ordered phase exhibits typical features of far-from-equilibrium ordered phases present in models that consider self-propulsion. For instance, our model shows \emph{giant number fluctuations}, that is a signature of fluctuating ordered active phases \cite{ton05}, here characterized by $\sigma(n)$, corresponding to the square root of the variance of the particles contained ($n = \rho l^2$ in average) in square boxes of linear size $l$. For our model, $\sigma(n)$ scales like $n^{\alpha}$ with $\alpha \simeq 0.9 > \frac{1}{2}$ [Fig.\ \ref{fig2}(b)]. This value is, however, larger than that known for the Vicsek \emph{et al.}\ model and its gene\-ralizations \cite{gin10}.

\begin{figure}
\includegraphics[width=0.48\textwidth]{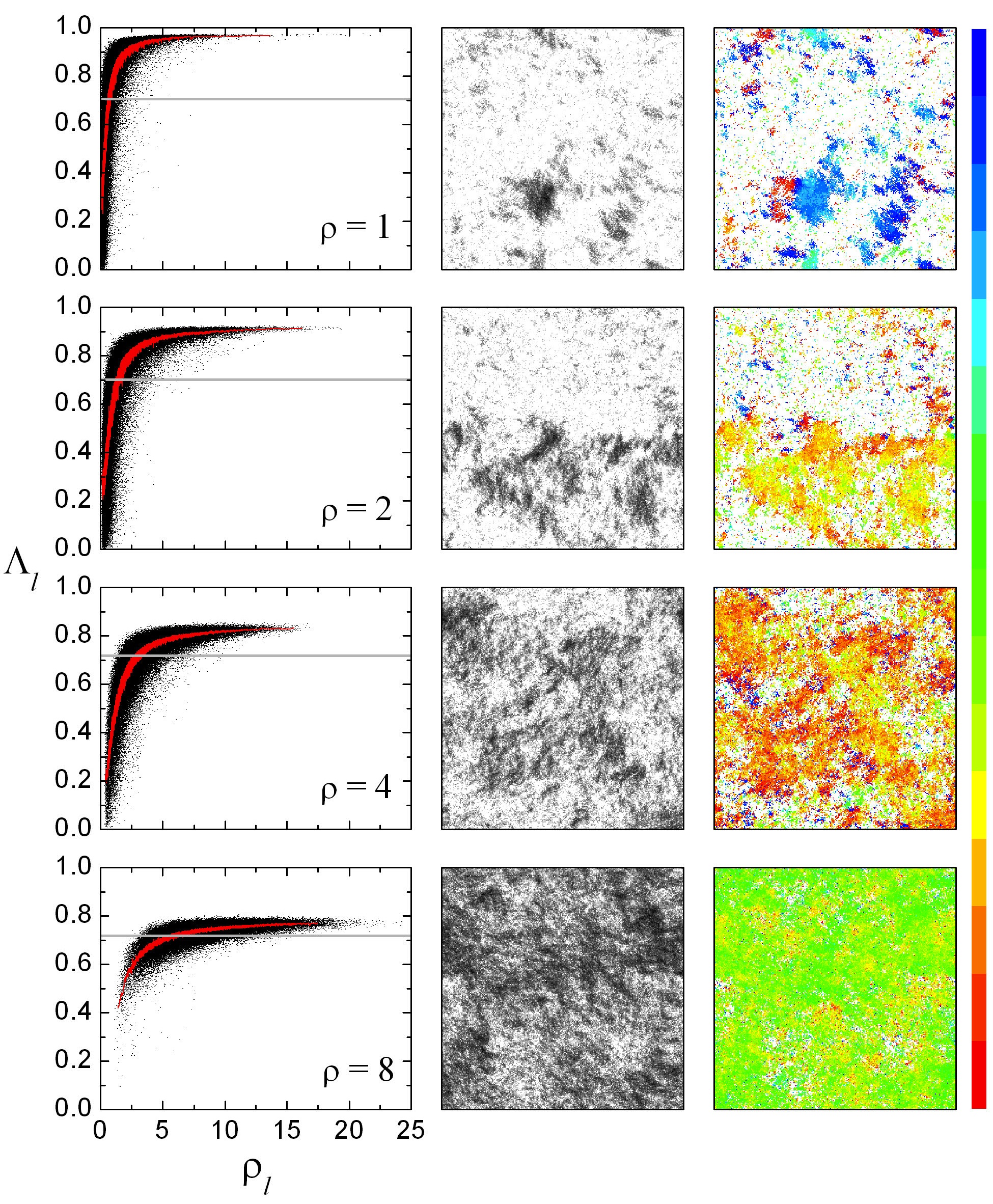}
\caption{(Color online) In the first column, scatter plots of local oder parameter $\Lambda_l$ versus local density $\rho_l$ computed in boxes of size $l=14$ for the same systems of figure \ref{fig1}(b) with $L=196$. The red lines passing through the middle of the scattered black dots are running averages in a local window in $\rho_l$. The horizontal grey line corresponds to the stationary order parameter $\langle\Lambda\rangle$. The middle and last columns show snapshots of the corresponding density and direction fields. For the middle column, the darker gray level represents higher particle density. The color key on the right goes from a direction 0 to $2\pi$, taken with respect to the horizontal positive axis, from bottom (red) to top (blue), respectively.}
\label{fig3}
\end{figure}

Surprisingly, even with self-propulsion left aside, our model exhibits true \emph{long-range order} as shown in Fig.\ \ref{fig2}(c), where the order parameter $\Lambda$ slowly decays (algebraically) with $L$ to a constant value $\Lambda_{\infty} > 0$ for all of the particle-density values considered here. Moreover, dilute systems with, \emph{e.g.}, $\rho = 1$, show what seems to be \emph{intermittent behavior} (also present in the original Vicsek \emph{et al.}\ model \cite{hue04}), where $P(\Lambda)$ displays an asymmetric Gaussian form with an exponential tail as shown in Fig.\ \ref{fig2}(e). Segregation decreases with $\rho$, while the distribution of $\Lambda$ approaches a Gaussian one, with $\Lambda$ showing less spikes of instantaneous disorder as a function of $t$ [Figs.\ \ref{fig2}(d) and \ref{fig2}(e)]. This is due to the fact that, as $\rho$ increases, the particles in the system become more homogeneously distributed over space [Fig.\ \ref{fig3}]. In fact, as $\rho \rightarrow \infty$ or if global coupling ($R/\mathcal{L} = 1$) is considered, this feature allows for the analytical treatment of the model where the instability of the disordered close-to-equilibrium state can be demonstrated, and the relation of our model to the \hbox{Kuramoto} model of synchronization is revealed. This case is discussed in details in \cite{sev14}.

Another feature displayed by ordered far-from-equilibrium phases, whenever the interactions depend on the distance, is the strong coupling between \emph{local density and local order} \cite{gre04, gin10}. As shown in Fig.\ \ref{fig3} for our model, this feature is more evident for dilute more-segregated systems, as the particles tend to concentrate in a few very dense and very ordered clusters. This behavior also decreases with $\rho$, due to the fact that the system becomes more homogeneous and correlated. See, for example, that the snapshots for the direction field (right co\-lumn of the figure) show less white spots as $\rho$ increases, meaning that, wherever there is color,  there are at least a few particles with their mean direction in correspondence to the color key.

\begin{figure}
\includegraphics[width=0.49\textwidth]{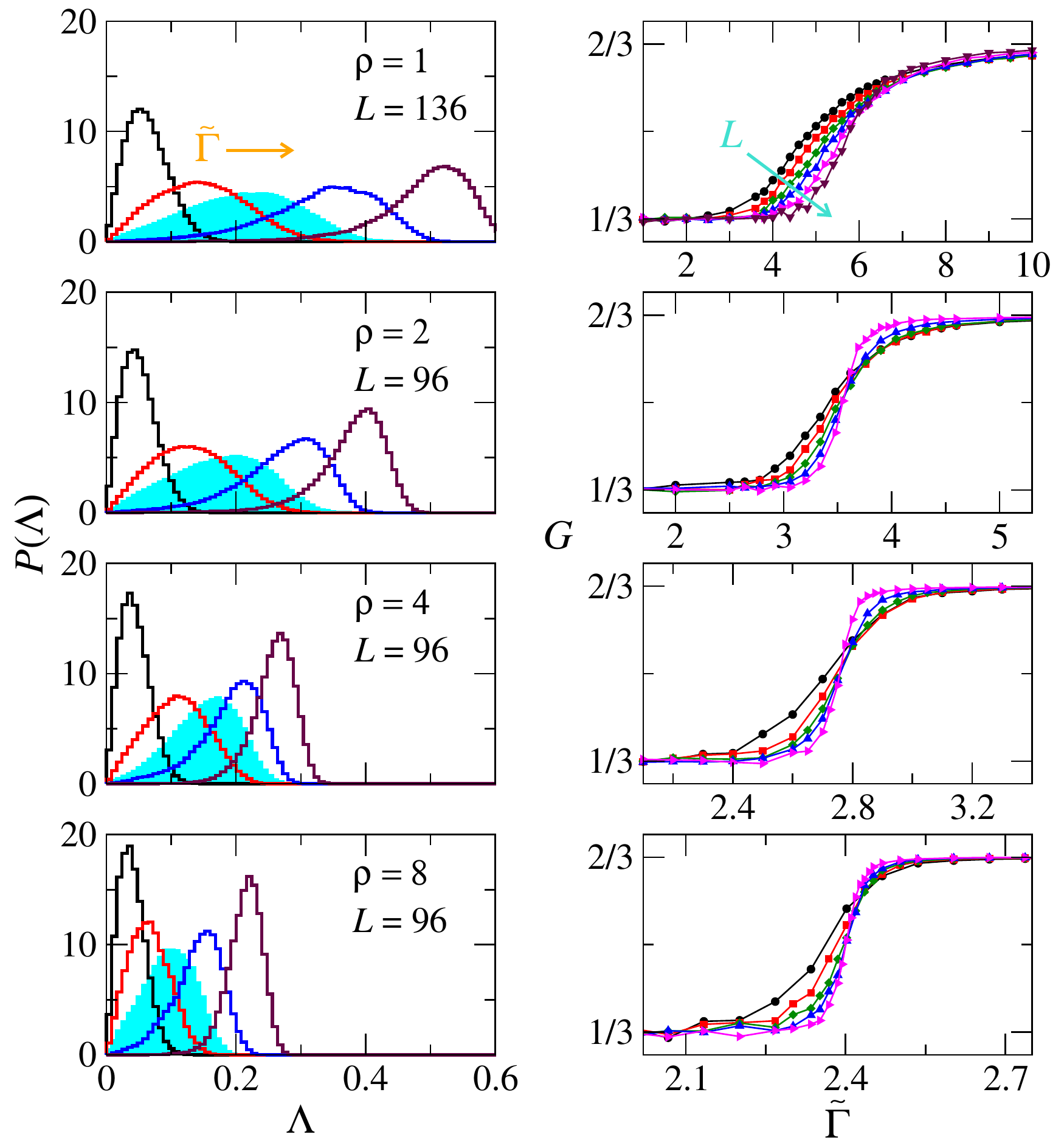}
\caption{(Color online) Left column, stationary $P(\Lambda)$ vs $\Lambda$ for different values of $\tilde{\Gamma}$ around the critical point. The filled curves denote the unimodal character of $P(\Lambda)$ in the transition from disordered to the ordered phases. On the right column, plots of the Binder cumulant $G$ vs $\tilde{\Gamma}$ for system with the corresponding density and $L=24, 34, 48, 68, 96, 136$. As can be appreciated, $G$ never becomes negative nor shows any discontinuity. Both of these features are signature of a second-order phase transition.}
\label{fig4}
\end{figure}

Results regarding the nature of the disorder-order phase transition displayed by our model are presented in Fig.\ \ref{fig4}. On the left column, the distribution $P(\Lambda)$ shows a single-peaked form around $\tilde{\Gamma}_c$ for the systems considered here. Additionally, on the right column, the behavior of the Binder cumulant \hbox{$G=1-\langle \Lambda^4\rangle/(3\langle \Lambda^2\rangle^2)$} vs $\tilde{\Gamma}$, for different system sizes, shows a smooth behavior without any signs of discontinuity. Even more, there are no signs of the emergence of density waves, present in some models that consider self-propulsion and display a discontinuous phase transition \cite{gre04}. All of these facts hint the continuous character of the phase transition in our model.

It is interesting to notice, though, that the phase transition of our model takes place for finite values of $\tilde{\Gamma}$ even in the limit cases when $\rho \rightarrow \infty$ or when global coupling is considered \cite{sev14}. In contrast, in the Vicsek \emph{et al.} model and some of its generalizations, in the same limits, the disordered phase only occurs at maximum noise (or, equivalently, in the infinite temperature limit) \cite{dos09}.

We must mention that, due to the stronger fluc\-tua\-tions of $\Lambda$ for dilute systems, the computing time required is larger than that required for systems with higher particle density, nonetheless, denser systems include more particles than dilute ones for the same values of $L$. For instance, our runs for the largest dilute systems required about $10^8$ integration steps. Even so, longer integration times with larger systems are required for the determination of quantities such as critical points and exponents for the different cases considered here. Moreover, larger systems require longer integration times due to the locality of the interactions. These calculations are beyond the computational resources available to us at the moment, however, we are woking on to study this in the future.

In summary, our results show that non-Hamiltonian interactions that do not preserve momentum (such as the alignment interaction typically used to model flocking behavior), in combination with underdamped particle dynamics, are able to drive the system from an close-to-equilibrium (disordered) phase to a far-from-equilibrium (ordered) one in a novel type of phase transition. Thus, we believe our model is sui\-ta\-ble for studying the passage from states where the entropy is maximized to statio\-na\-ry phases where entropy is produced. We are currently pursuing this line of investigation. In addition, under such circumstances, we have also demonstrated that self-propulsion is an unnecessary feature for systems to develop long-range order out of local interactions, even in two dimensions, as typically considered by some other models for flocking.

Beyond the obvious implications in the study of flocking phenomena, we believe our results are relevant in the more general context of the study of phase transitions in equi\-li\-brium or out-of-equilibrium statistical mechanics.

\begin{acknowledgments}

The authors gratefully acknowledge the computing time granted on the super\-com\-pu\-ters MIZTLI (DGTIC-UNAM), THUBAT-KAAL (CNS-IPICyT) and, through the project ``Cosmolog\'{\i}a y as\-tro\-f\'{\i}\-si\-ca relativista: objetos compactos y materia obs\-cu\-ra'', on \hbox{XIUHCOATL} (\hbox{CINVESTAV}). V.D.\ acknow\-led\-ges support from the grant PROMEP/103.5/10/7296 and from CONACyT. F.J.S.\ acknowledges support from the grant PAPIIT-IN113114.

\end{acknowledgments}

\end{document}